\documentclass[prb,preprint]{revtex4-1} 

\usepackage{amsmath}  
\usepackage{amsfonts} 
\usepackage{graphicx} 
\usepackage{wrapfig}
\usepackage{subcaption}  
\usepackage{natbib}
\begin{document}


\title{A Simple Miniaturized Stellar Interferometer: Measuring the Angular Diameter of Venus with a Masked Telescope}

\author{Eric J. Friedman}
\email{ejficsi@gmail.com} 
\affiliation{International Computer Science Institutem (Retired), Berkeley, CA} 
\author{Michelle Goman}
\email{goman@sonoma.edu} 
\affiliation{Sonoma State University, Rohnert Park, CA}


\date{\today}

\begin{abstract}

We describe a simple, visually compelling demonstration/laboratory of astronomical interferometry using a standard planetary telescope and a thin piece of opaque plastic punched with two small holes. By observing the disappearance of interference fringes as the hole spacing is increased, students can directly estimate the angular diameter of Venus. The demonstration is a tabletop analogue of the landmark 1921 Michelson–Pease measurement of Betelgeuse \citep{michelson1921} and scales naturally from that measurement: where Michelson and Pease required a 6.1 m baseline to resolve a disk 47 mas (milli-arcsecond) across, a 15 mm baseline suffices for Venus, when it is approximately 10-15$''$ (arc seconds). The lab works reliably in ordinary atmospheric seeing conditions, and a side-by-side comparison of Venus (or Jupiter) with a bright star such as Sirius provides an immediate visual demonstration of the difference between extended and point sources. It works equally well as a laboratory at a variety of difficulty levels. Furthermore, it does not require advanced knowledge or skills with optical telescopes. While the underlying ideas are well known, we are unaware of any source that presents them in  an accessible, unified form.

\end{abstract}

\maketitle 



\section{Introduction} 

Interferometry is one of the most powerful techniques in modern astronomy. Its significance stems from a fundamental limitation of conventional telescopes: the angular resolution of a telescope is essentially set by its aperture diameter. The key insight of interferometry is that two smaller telescopes separated by a distance $d$, the baseline, can achieve angular resolution similar to a single aperture of diameter $d$. This makes it possible to achieve resolutions that are orders of magnitude finer than those of any single telescope.

\begin{figure}[htbp]
  \centering
  \begin{subfigure}[b]{0.40\linewidth}
    \includegraphics[width=\linewidth]{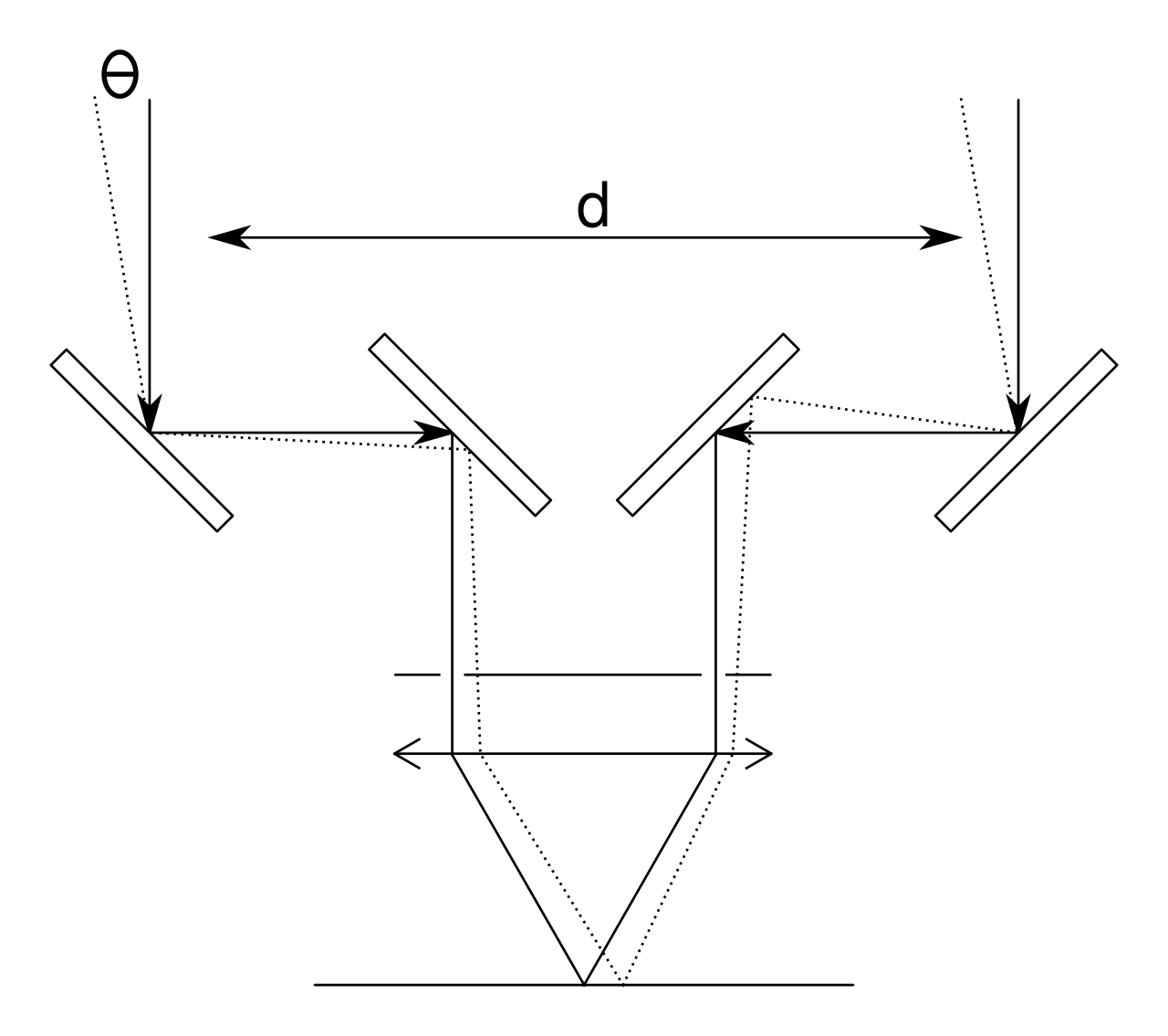}
    \caption*{(a)}
  \end{subfigure}
  \hfill
  \begin{subfigure}[b]{0.40\linewidth}
    \includegraphics[width=\linewidth]{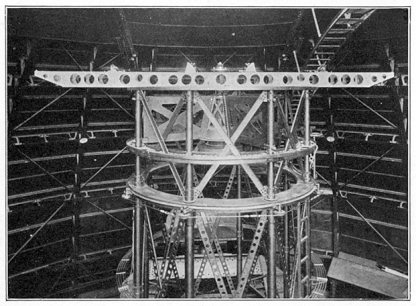}
    \caption*{(b)}
  \end{subfigure}

  \vspace{0.5em}

  \begin{subfigure}[b]{0.40\linewidth}
    \includegraphics[width=\linewidth]{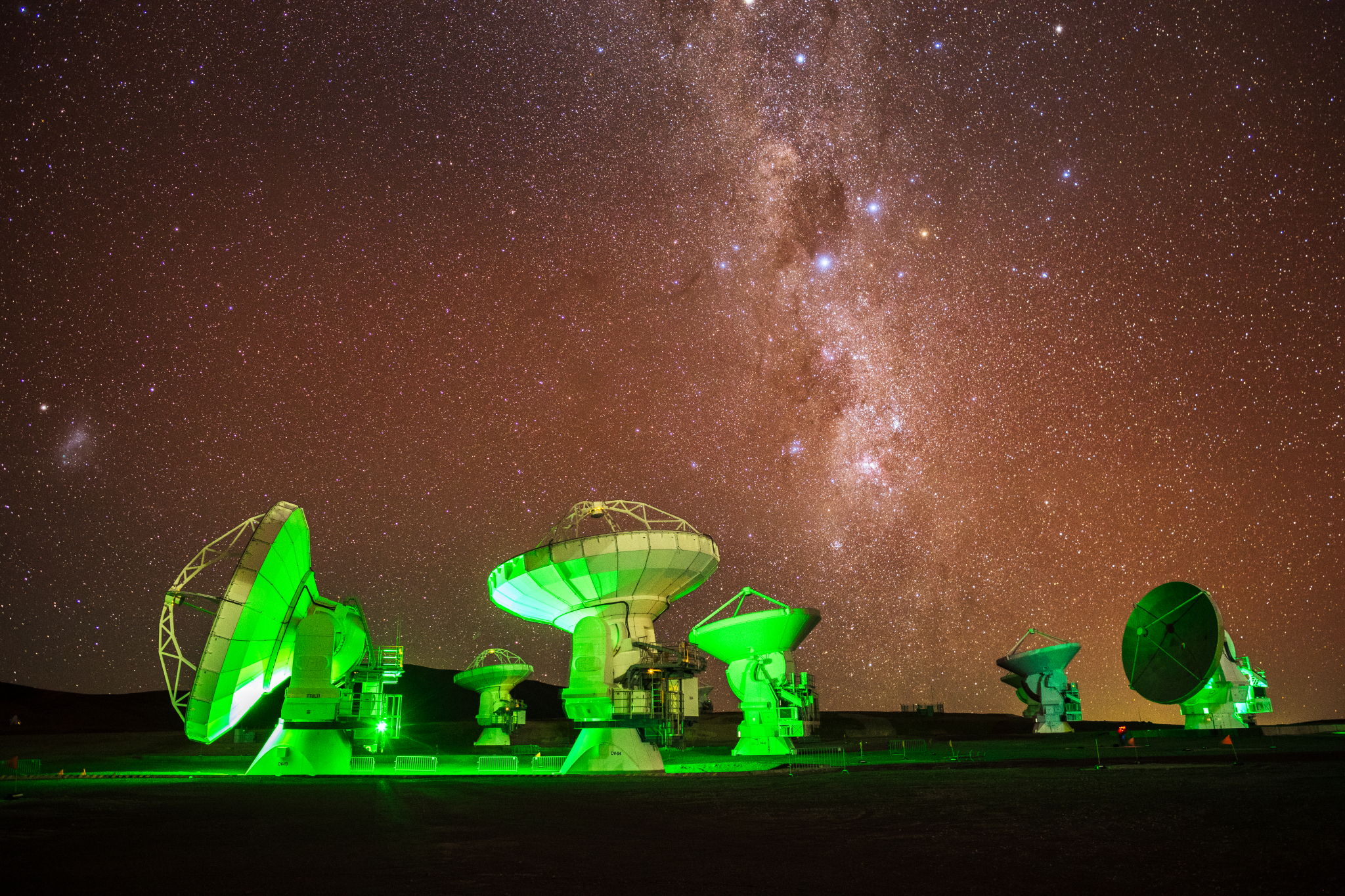}
    \caption*{(c)}
  \end{subfigure}
  \hfill
  \begin{subfigure}[b]{0.40\linewidth}
    \includegraphics[width=\linewidth]{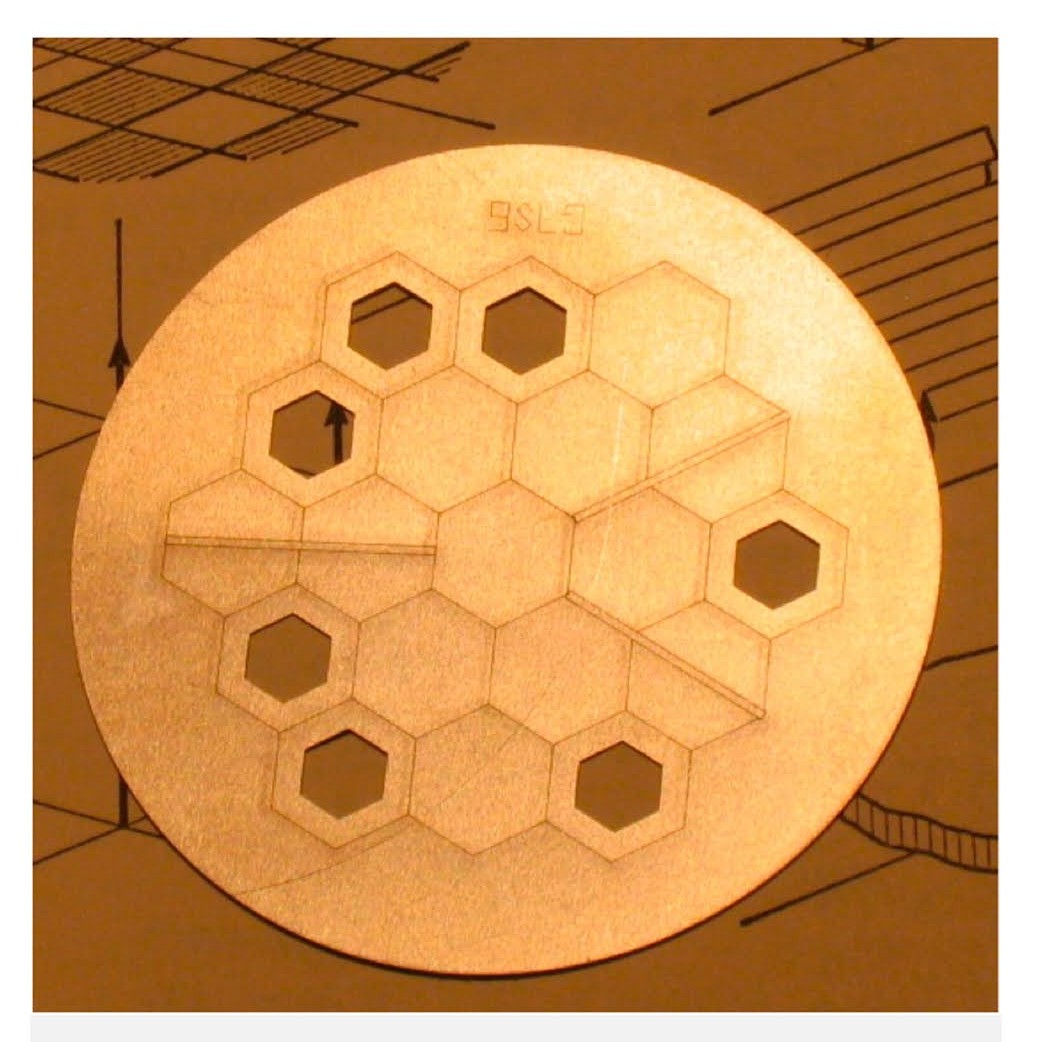}
    \caption*{(d)}
  \end{subfigure}

  \caption{\raggedright (a) Schematic of the Michelson stellar interferometer. (b) The Michelson interferometer mounted on the frame of the Hooker Telescope, 1920. (c) The ALMA radio interferometer. (d) A prototype of the non-redundant mask for the James Webb Space Telescope.}
  \label{fig:inters}
\end{figure}

The technique is now indispensable in astronomy. In the optical and near-infrared, long-baseline interferometers such as the CHARA Array (baseline up to 330 m) \citep{ten2016update} and the VLTI (baseline up to 200 m) \citep{glindemann2000vlt} routinely measure stellar diameters, orbits of binary stars, and the structure of circumstellar disks. Aperture masking  is used on instruments including the James Webb Space Telescope's NIRISS instrument  \citep{sivaramakrishnan2012non}. (Figure~\ref{fig:inters} (d).) In radio astronomy, Very Long Baseline Interferometry links antennas scattered across entire continents, achieving baselines of up to 10,000 km or more and sub-mas (milli-arcsecond) resolution. (Figure~\ref{fig:inters} (c).)

Despite its importance, stellar interferometry is challenging to demonstrate convincingly. Most optics labs use laser sources and bench-top mirrors, which, while precise, offer little intuition for how astronomers measure stars\citep{libbrecht2015basic,pryor1959measuring,illarramendi2014daylight}. Laboratory applications that use radio waves to observe the Sun are more authentic, but mechanically and analytically complex \citep{koda2016michelson} and require sophisticated electronics and signal processing. Labs that resolve binary stars also require substantial data acquisition and processing \citep{argyle2012observing}. 

In this article, we describe a lab that is both visually immediate and physically authentic: using a masked telescope to observe Venus, students can directly see fringes appear and vanish as the hole separation changes, and from this directly infer its angular size. A simple comparison of Jupiter and a bright star such as Sirius can be used to provide an immediate visual demonstration of the difference between extended and point sources. Although we have found a discussion of this in the literature\cite{GlindemannSpatialInterferometry} we have not been able to find any actual implementations of this and many simple approaches we have experimented with turned out to be extremely fragile in practice. The main goal of this paper is to present a simple robust method that we have developed. This allows one to reliably demonstrate the basic techniques of stellar interferometry in a single evening lab or even public demonstration.

Note that the lab does not require advanced knowledge or skills with optical telescopes. If  a student can find Venus in a reasonably high magnification eyepiece they can do the lab  without knowing anything about focal lengths, magnification computations etc.  The other tools are similarly low tech for the basic analysis, solid thin opaque sheets of plastic or paper and  a drill to cut clean holes. In fact, we have even had success with sheets of printer paper and holes poked with a ballpoint pen!

This project began with a search for a simple intuitive demonstration of interferometry. While all the key ideas used in this lab are well known, we were surprised that we could find no simple examples in the literature.  In fact we could not find any nonprofessional uses of interferometry aside from a brief suggestion for its use in binary star systems \citep{argyle2012observing} and a sophisticated radio astronomy lab \citep{koda2016michelson}. Surprisingly, more complicated approaches that we tried, failed or had little benefit; while our very simple approach is both intuitive and convincing and allows for interesting and more sophisticated extensions.

\section{Historical Background}

The measurement that inspired this lab is the 1921 determination of the angular diameter of Betelgeuse ($\alpha$ Orionis) by Albert Michelson and Francis Pease \citep{michelson1921}. Working at the Mount Wilson Observatory, they mounted a 20-foot steel beam on the 100-inch  Hooker telescope. Four flat mirrors directed starlight into the telescope's light path, creating interference fringes. When the outer mirror separation exceeded $\sim 3$ m, the fringes disappeared, indicating that Betelgeuse's disk was spatially resolved in the interferometer.  (Figure~\ref{fig:inters} (a) and (b).) From this they inferred an angular diameter of approximately 47 mas, a value in good agreement with modern measurements of ~55 mas.

The lab described here is a miniaturization of these classic measurements. In place of Michelson and Pease's 6.1 m baseline we use a baseline of about 20 mm, and in place of Betelgeuse ($\sim 47$ mas) we observe Venus, at a point where its angular diameter is approximately 10$''$ — roughly 200 times larger than Betelgeuse but also $\sim 100$ times brighter, allowing amateur optics to be used.

\section{Theory}

In this section we review the theory  behind our analysis as it is important for our design. We follow the classic reference by Born and Wolf\citep{born1999}.  The reader might also refer to Koda et al.\cite{koda2016michelson} for a direct computation and a more formal approach to this analysis.

\subsection{Resolution of a Single Circular Aperture}

When a plane wave from a distant point source passes through a circular 
aperture of diameter $d$ and is focused by a lens, 
diffraction produces 
a characteristic intensity pattern in the focal plane — the Airy pattern. 
The angular radius to its first dark ring is
\begin{equation}
  \theta_{\rm Airy} = 1.22\,\frac{\lambda}{d}.
  \label{eq:rayleigh}
\end{equation}
This is the Rayleigh criterion: two incoherent point sources separated by
$\theta_{\rm Airy}$ are just resolved.
\begin{figure}[htbp]
  \centering
  \begin{subfigure}[b]{0.45\textwidth}
    \includegraphics[width=\linewidth]{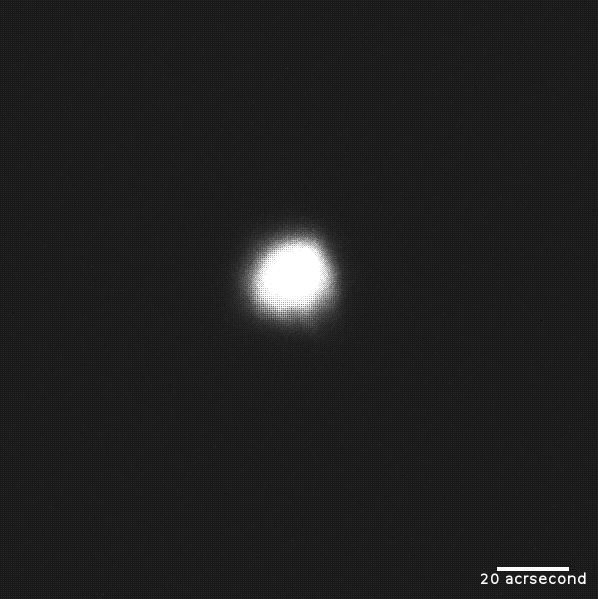}
    \caption*{(a)}
  \end{subfigure}
  \hfill
  \begin{subfigure}[b]{0.45\textwidth}
    \includegraphics[width=\linewidth]{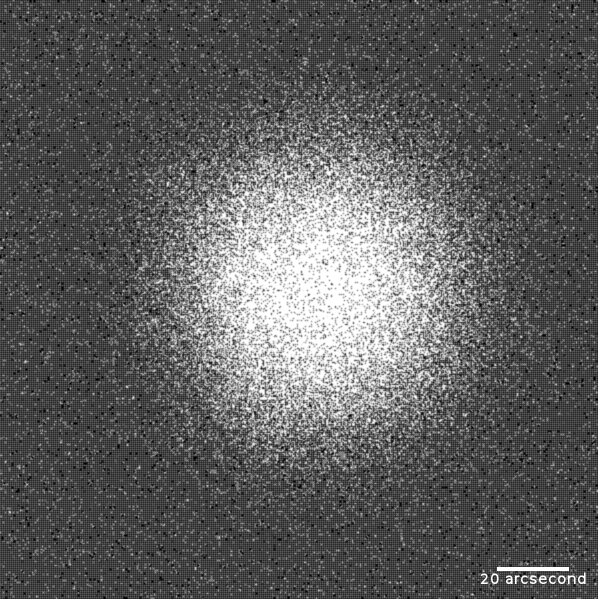}
    \caption*{(b)}
  \end{subfigure}

  {
  \caption{\raggedright (a) The (mostly) resolved disk of Venus $\sim 20$$''$ with 150 mm aperture compared to (b) Venus through a 2 mm aperture. Note that the Airy pattern appears to be about 6--7 times the size of the resolved disk. The observed Airy disk has a radius of approximately 60$''$, close to the theoretical value of 69$''$, although hard to quantify precisely. (Taken May 2026 when the diameter of Venus was $\sim 12$$''$ with extremely bad seeing, leading to significant lack of resolution for Venus in the 150 mm aperture.)}
  \label{fig:venus_airy}}
  
\end{figure}

For a  telescope aperture $d = 150\,\text{mm}$ at visible light with
$\lambda = 550\,\text{nm}$,
\begin{equation}
  \theta_{\rm Airy}(d)  \approx 2'',
\end{equation}
sufficient to resolve the disks of Venus, Mars, and Mercury, but not Betelgeuse. (See figure \ref{fig:venus_airy}.)

Note, that all of our measurements are completed with unfiltered white light; for the computations we choose a specific wavelength to simplify the analysis.  Again, this is meant to simplify the lab for the students.

\paragraph{Choice of hole diameter $h=2$ mm.}

In our lab the full aperture is replaced by a mask with two small circular holes, each of diameter $h$. For $h=2$ mm a simple calculation shows that this reduces the light collected by a factor of about 3000 or about 8 astronomical magnitudes of brightness. In addition the for $h=2$ mm the Airy disk is about $69''$, so is large enough to observe the fringes that cover it.

Venus at its brightest has a brightness of about  -4.6 mag and therefore appears as an
equivalent $\approx +3\,\text{mag}$ source through the mask which is comfortably
visible.  This 8 mag reduction is
the primary reason $h = 2\,\text{mm}$ is chosen: smaller holes would reduce fringe signal further, while appreciably larger holes would reduce the size of the fringes and the Airy disk as we discuss below. 

\subsection{Interference Fringes from a Two-Hole Mask}

Consider a mask with two holes of diameter $h$ whose centers are
separated by $d > h$.  A plane wave falling on this mask produces
two mutually coherent secondary sources.  In the focal plane the
intensity is the product of two factors: the broad Airy-disk envelope
of each individual hole and a cosine-squared fringe pattern.  The theoretical
angular spacing between adjacent bright fringes is
\begin{equation}
  \Delta\theta_{\rm fringe} = \frac{\lambda}{d},
  \label{eq:fringe_spacing}
\end{equation}
The theoretical number of fringe maxima visible within the central envelope
lobe is  
\begin{equation}
  N_{\rm fringes} \approx \frac{2\,\theta_{\rm Airy}}{\Delta\theta_{\rm fringe}}
                        = 2.44\,\frac{d}{h}.
  \label{eq:N_fringes}
\end{equation}

As a numerical example, for $h = 2\,\text{mm}$, $d = 6\,\text{mm}$,
and $\lambda = 550\,\text{nm}$:
\begin{align}
  \Delta\theta_{\rm fringe} &= \frac{550\times10^{-9}}{6\times10^{-3}}
    \approx 19'', \\
     \quad N_{\rm fringes}\approx 7.
\end{align}
The fringe spacing  exceeds typical atmospheric seeing ($1''$–$5''$),
so individual fringes can be resolved through an eyepiece or camera. However, note that this is a theoretical upper bound and in practice significantly fewer fringes are typically discernible. (See Figure \ref{fig:venus_fringes} (a).)

\subsection{Effect of Finite Source Size: Fringe Observability}

A spatially extended source may be treated as a continuous ensemble of 
incoherent point sources, each producing its own fringe pattern shifted 
in phase proportional to its angular offset from the optical axis.  As the hole separation $d$ increases, the fringe patterns from 
opposite edges of the disk arrive increasingly out of phase, and the 
fringes progressively wash out. At a critical separation the bright 
fringes from one half of the disk exactly fill in the dark fringes from 
the other half, leaving uniform illumination with no visible contrast. Note that this is a theoretical bound and differs from the previous section; in practice fringes typically are not discernible at smaller separations.

One increases $d$ by swapping masks until the fringes first 
vanish; this separation is called the null baseline $d_{\rm null}$.
For a uniformly bright circular disk of angular diameter $\alpha$, the 
van~Cittert--Zernike theorem  gives
\begin{equation}
  d_{\rm null} = \frac{1.22\,\lambda}{\alpha}.
  \label{eq:dnull}
\end{equation}

\begin{figure}[htbp]
  \centering
  \begin{subfigure}[b]{0.45\linewidth}
    \includegraphics[width=\linewidth]{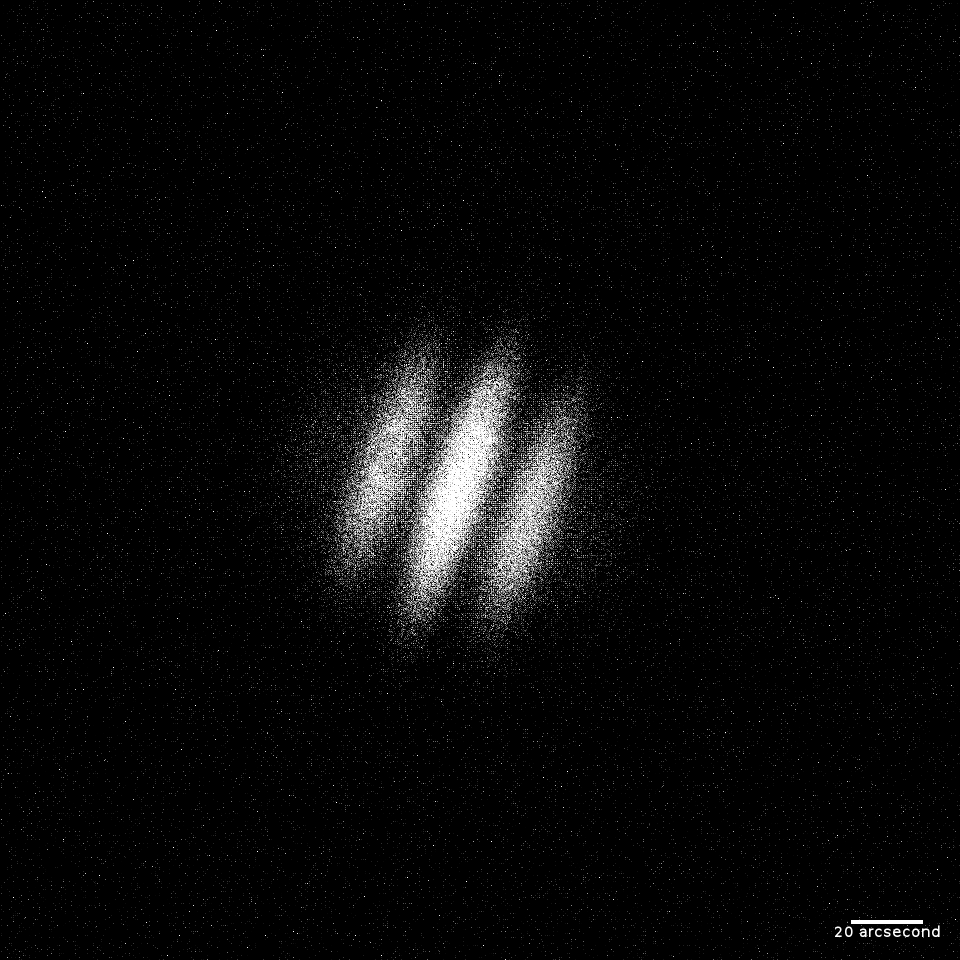}
    \caption*{(a)}
  \end{subfigure}
  \hfill
  \begin{subfigure}[b]{0.45\linewidth}
    \includegraphics[width=\linewidth]{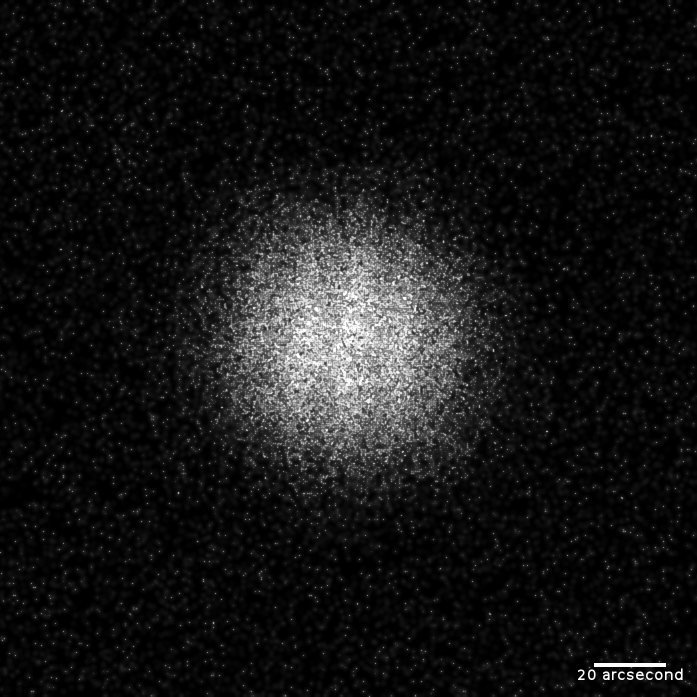}
    \caption*{(b)}
  \end{subfigure}

  \vspace{0.5em}

  \begin{subfigure}[b]{0.45\linewidth}
    \includegraphics[width=\linewidth]{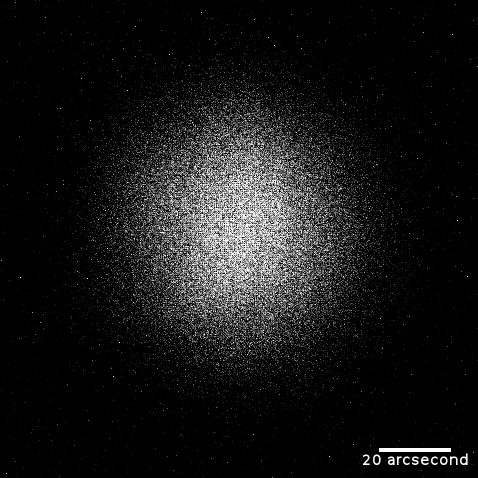}
    \caption*{(c)}
  \end{subfigure}
  \hfill
  \begin{subfigure}[b]{0.45\linewidth}
    \includegraphics[width=\linewidth]{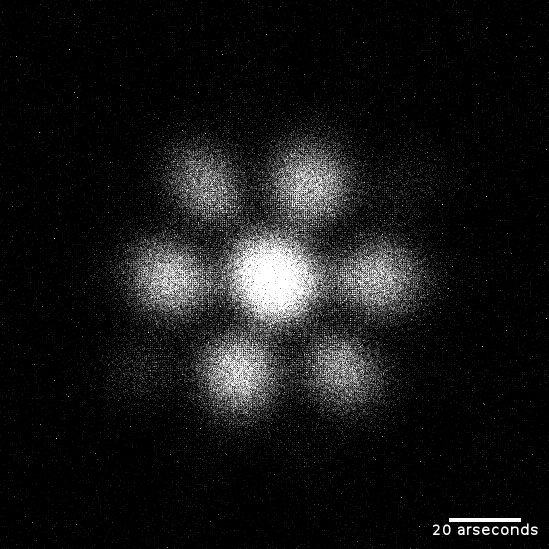}
    \caption*{(d)}
  \end{subfigure}

  {
  \caption{\raggedright Images of Venus (May 2026, about 13$''$ in size) through a two-hole mask with 2 mm holes. Note that the outer fringes are not seen as they are dimmer and smaller than the central fringes. (a) $d = 6\,\text{mm}$: the diameter is less than 18$''$ and the fringes are clear. (b) $d = 10\,\text{mm}$: near the first null, fringes are somewhat visible, corresponding to 11$''$. (c) $d = 14\,\text{mm}$: complete lack of fringes, corresponding to 8$''$. Thus, the measured size is approximately 11$''$, reasonably close to the true value. (d) The interference pattern for a triangle of 2\,mm holes, 6\,mm on each side, creating a hexagonal pattern.}
  \label{fig:venus_fringes}
  }
\end{figure}

For Venus at its
smallest ($\alpha \approx 10''$) gives $d_{\rm null} \approx 13\,\text{mm}$, 
while Jupiter near its minimum apparent size ($\alpha \approx 30''$) 
only gives $d_{\rm null} \approx 4.6\,\text{mm}$.
(See Figure \ref{fig:venus_fringes} (a,b,c).)

\subsection{Choice of Baseline Range: Venus vs.\ Other Planets}
\label{sec:baseline_choice}

To reiterate, Venus at its smallest is sufficiently bright and small to have a constructable mask that can be used to estimate its size to reasonable accuracy. Jupiter at is smallest is quite marginal and the other planets are extremely challenging, although as we mention below the use of a high quality astronomical camera mitigates some of these issues and lead to interesting extensions of this as a student lab.

\subsection{Relation Between Angular and Physical Diameter}

We remind the instructor that the measurement described above yields the angular diameter $\alpha$,
not the physical diameter, a technicality that may confuse beginning students.  Converting between the two requires knowledge
of the planet's distance $r$ and  determining this  is  typically covered in basic astronomy classes.

\section{The lab}

\subsection{Equipment}

{\bf Telescope}:  The type of telescope is quite flexible. Any telescope that can view Venus or Jupiter steadily at high magnification will be sufficient. We have used both a 150 mm  reflector with  a planetary camera\footnote{All images were taken with a Celestron NexStar 6se telescope and a ZWO ASI676MC camera.} and a 114 mm refractor with a high magnification lens with success. We emphasize that a simple quality telescope on a solid mount is sufficient as the measurement can be done in a few seconds, so no automated tracking or expensive mounts are required.

\begin{figure}[htbp]
  \centering
  \begin{subfigure}[b]{0.48\textwidth}
    \includegraphics[width=\linewidth]{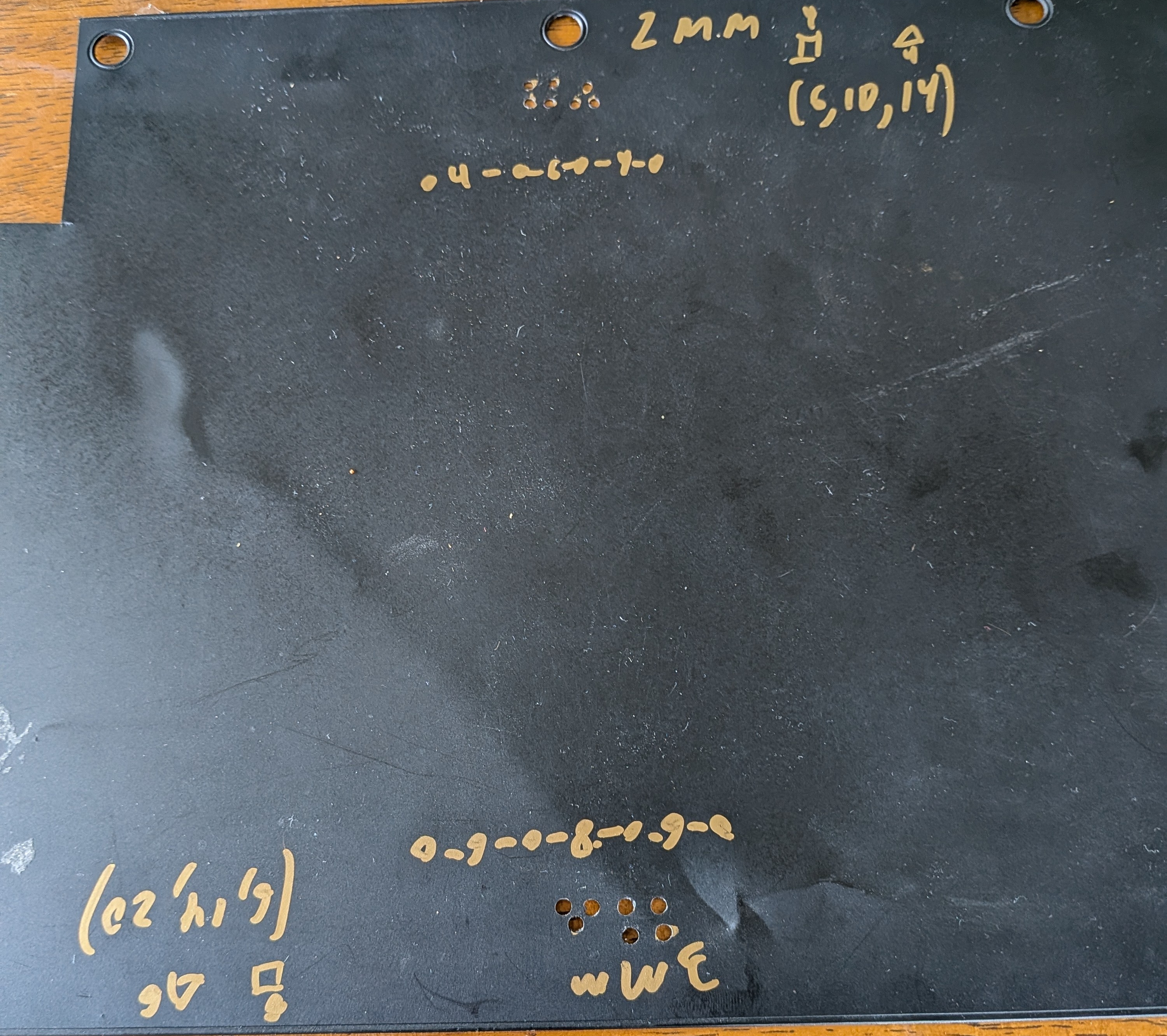}
    \caption*{(a)}
  \end{subfigure}
  \hfill
  \begin{subfigure}[b]{0.48\textwidth}
    \includegraphics[width=\linewidth]{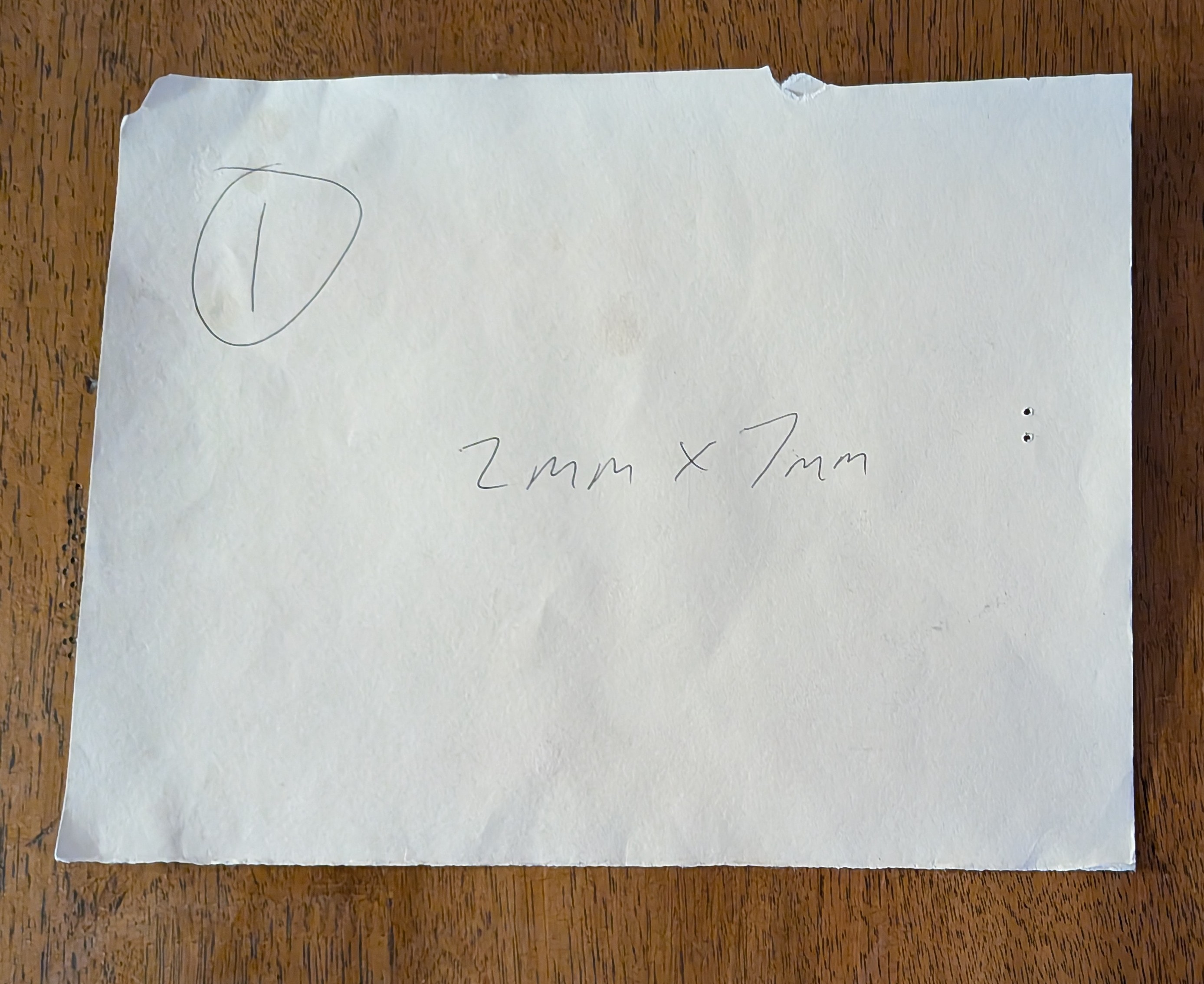}
    \caption*{(b)}
  \end{subfigure}

  {
  \caption{\raggedright (a) Plastic mask with multiple geometries. Note that the holes are away from the center to avoid the blind spot of a reflector telescope. Unused holes are covered with masking tape as needed. (b) Prototype mask made from a manila folder.}
  \label{fig:masks}
  }
\end{figure}

{\bf Mask}: Any opaque, thin but stiff material can be used. We have used dark plastic sheets as well as manila folders. It is important to make clean holes.  We used a power drill with a 5/64 bit; place the paper on a piece of wood and drill carefully, as the holes need to be clean. The location of the holes, other than the spacing, is flexible. For a refractor scope you can put the holes toward the center of the objective lens while for many reflector scopes you want the holes near the edge to avoid the central blind spot. Choose an assortment of distances from 5 mm and 20 mm. Using simple inexpensive materials allows students to create a wide variety of masks to experiment with. (Figure~\ref{fig:masks}.) See next section.

\subsection{Experimental Technique}

 Center the object in the eyepiece (or camera) and then quickly (before the target can move) place the mask over the aperture and rate the fringes in terms of clarity. Vary the masks and find the mask size at which the fringes first disappear or minimize. A simple  check consists of rotating the mask and having the observer identify the direction of the fringes (which should be perpendicular to the line between the holes). Recall that the image dims considerably with the mask.

Additional tasks include counting the number of fringes and comparing that to the theory. 

\subsection{Targets}

As discussed above Venus, when small, bright and separated from the sun enough to measure its angular size, is the ideal target. This occurs somewhat before greatest elongation on the way to superior conjunction.  Unfortunately this only occurs about every 19 months.

Note that other planets are extremely difficult to measure without advanced measurement techniques. First, the light is limited by the holes in the mask so larger aperture telescopes do not improve this. Second, in theory one could use smaller holes closer together to measure Jupiter, but this requires both precision fabrication and longer exposures necessitating tracking and imaging equipment.

Alternatively, one can simply view a bright star (such as Sirius or Arcturus) to see that fringes are strong for any reasonable spacing. In this case counting fringes and comparing to theory makes an interesting lab.  One can then compare this to Jupiter, which  for feasible mask sizes shows negligible fringing.  This provides a decent lower bound of about 10$''$ for Jupiter (30-50$''$) and a very poor upper bound which is at best 0.1$''$ for the angular size of a star (less than 0.05$''$). 

\subsection{Analysis and Extensions}

The analysis can be as simple as deciding whether fringes exist or an extremely careful analysis of fringe size and strength as a function of mask size.  One could give students a great deal of leeway to try various combinations of hole sizes and mask spacings, given the ease of constructing masks. Glindemann provides a detailed analysis for more advanced measurements.\cite{GlindemannSpatialInterferometry}

We emphasize the possibility of a simple demonstration for beginning students where you show that stars have fringes with a mask, but planets, such as Jupiter, do not. Also, we find that the basic aperture effect, that a small aperture makes the object significantly larger is surprising and instructive for beginners. We have even successfully demonstrated these effects at a public star night.

Alternatively one can vary masks. For example using slits instead of holes increases light throughput and  with careful design allows one to measure the size of Venus when it is larger than 12$''$, Jupiter and possibly even Mars.

Using multiple holes is also interesting. For example two rows of holes return similar results to a pair of holes but allow more light. Other geometries of holes lead to interesting fringe patterns, e.g. Figure \ref{fig:venus_fringes} (d). One could even try non-redundant masks where a selection of holes are constructed to create a variety of distances which allows a single mask to detect a variety of object sizes, as used in the James Webb space telescope \citep{sivaramakrishnan2012non}. (Figure~\ref{fig:inters} (d).)

In addition, one could  increase the sophistication and power of the lab by adding cameras with long exposure times, allowing for smaller hole sizes and spacing in order to measure Jupiter or Saturn, or larger holes and spacings to measure Mars or perhaps even Io or Mercury. This setup also can be used to analyze double stars \cite{argyle2012observing}.

\subsection{Comments}

Surprisingly, atmospheric seeing is not particularly significant due to the small apertures used, a well known advantage of interferometry.

For convenience, one does not need to wait for a fully dark sky. We have successfully observed the effect at civil sunset.

Perhaps the greatest difficulty for students is keeping the planet centered at high magnification.  Having one person place and remove the mask while the other keeps it aligned may require some practice.

\section{References}

\bibliography{inter}

\end{document}